\documentclass[a4paper,12pt]{article}
\usepackage{epsfig}
\usepackage{color}
\usepackage[numbers,sort&compress]{natbib}

\newlength{\dinwidth}
\newlength{\dinmargin}
\usepackage{fancyhdr}

\begin{document}

\title{Analysis of $B_s\to \phi\mu^+\mu^-$ decay within supersymmetry}

\author{
\quad Yuan-Guo  Xu, \quad  Li-Hai Zhou,
 \quad Bing-Zhong Li,
 \quad Ru-Min Wang\thanks{E-mail: ruminwang@gmail.com}\\
{\scriptsize { \it College of Physics and Electronic Engineering,
Xinyang Normal University,
 Xinyang, Henan 464000, China
}}
\\
 {\scriptsize  }
 }

\maketitle \thispagestyle{fancy} \fancyhead[RO]{Submitted to
`Chinese Physics C'}

\begin{abstract}
 Motivated by the first measurement on
$\mathcal{B}(B_s\to\phi \mu^+\mu^-)$ by the CDF Collaboration, we
study the supersymmetric effects in semi-leptonic $B_s\to
\phi\mu^+\mu^-$ decay. In our evaluations, we analyze the
dependences of the dimuon invariant mass spectrum and the
forward-backward asymmetry on relevant supersymmetric couplings in
the MSSM with and without R-parity. The analyses show the new
experimental upper limits of $\mathcal{B}(B_s\to\mu^+\mu^-)$ from
the LHCb Collaboration could further improve the bounds on sneutrino
exchange  couplings and $(\delta^u_{LL})_{23}$ as well as
$(\delta^d_{LL,RR})_{23}$ mass insertion couplings. In addition,
within the allowed ranges of relevant couplings under the
constraints from $\mathcal{B}(B_s\to\phi \mu^+\mu^-)$,
$\mathcal{B}(B\to K^{(*)}\mu^+\mu^-)$ and $\mathcal{B}(B_s\to
\mu^+\mu^-)$, the dimuon forward-backward asymmetry and the
differential dimuon forward-backward asymmetry of $B_s\to\phi
\mu^+\mu^-$ are highly sensitive to the squark exchange contribution
and the $(\delta^u_{LL})_{23}$ mass insertion contribution. The
results obtained in this work will be very useful in searching
supersymmetric signal at the LHC.
\end{abstract}

\section{Introduction}
Flavor changing neutral current (FCNC) processes can occur via
penguin or box diagrams in the standard model (SM) and are very
sensitive to the gauge structure as well as various extensions of
the SM. So they can provide useful information on the parameters of
the SM and test its
 predictions. Meanwhile, they can offer a valuable possibility of an
 indirect search of new physics (NP).
The transition $b\to s\mu^+\mu^-$, for instance, is a FCNC process,
present in the decays $B\to K\mu^+\mu^-$, $B\to K^{*} \mu^+\mu^-$,
$B_s\to\phi \mu^+\mu^-$ and $B_s\to \mu^+\mu^-$. The rates for these
decays could be changed by NP contributions, and this would
consequently alter the dimuon invariant mass spectra and the
forward-backward asymmetries for these semi-leptonic decays from the
SM predictions.

Recently, the first measurements of the branching ratio
\cite{Aaltonen:2011cn} and the dimuon invariant mass spectrum
\cite{Aaltonen:2011qs} of $B_s\to\phi \mu^+\mu^-$ have been reported
by CDF Collaboration. And its branching ratio is
\cite{Aaltonen:2011qs}
\begin{eqnarray}
 \mathcal{B}(B_s\to \phi\mu^+\mu^-) =
 (1.47\pm0.24\pm0.46)\times10^{-6},
\end{eqnarray}
which is quite consistent with its SM prediction
$\mathcal{B}^{SM}(B_s\to
\phi\mu^+\mu^-)=(1.48^{+2.06}_{-0.46})\times10^{-6}$.
Moreover, the upper limit of $\mathcal{B}(B_s\to \mu^+\mu^-)$ has
been significantly improved by the CDF, CMS and LHCb Collaborations
\cite{Aaltonen:2011fi,Chatrchyan:2012rg,arXiv:1112.0511,LHCBdata,CMSLHCB}.
The lowest published limit from the LHCb Collaborations at 95\%
confidence level (CL) is \cite{Aaij:2012ac}
  \begin{eqnarray}
   \mathcal{B}(B_s\to \mu^+\mu^-)<4.5\times10^{-9}.
\end{eqnarray}
These observables are important to test the SM and constrain
contributions of the possible NP models. Thus, these processes  have
attracted much attention (for instance, Refs.
\cite{Lunghi:2010tr,Yilmaz:2008pa,Zebarjad:2008un,Chang:2011jka,Li:2011yn,Dutta:2011bk,Beskidt:2011qf}).

In this paper, following closely the analyses of Ref.
\cite{Wang:2011aa}, we will study R-parity violating (RPV)
supersymmetric  effects and  the R-parity conserving (RPC) mass
insertion (MI) supersymmetric effects on the observables of $B_s\to
\phi\mu^+\mu^-$ decay from the new experimental data in the minimal
supersymmetric standard model (MSSM).
Using the experimental limits on $\mathcal{B}(B_s\to \mu^+\mu^-)$
from LHCb \cite{Aaij:2012ac}, $\mathcal{B}(B\to K^{(*)}\mu^+\mu^-)$
from Particle Data Group \cite{PDG} as well as $\mathcal{B}(B_s\to
\phi\mu^+\mu^-)$ from CDF \cite{Aaltonen:2011qs}, we will constrain
the relevant new couplings and examine the supersymmetric effects on
the branching ratio, the dimuon invariant mass spectrum, the forward
backward asymmetry and the differential forward backward asymmetry
of $B_s\to \phi\mu^+\mu^-$ decay.

The paper is arranged as follows. In Section 2,  we briefly
introduce the theoretical framework for $B_s\to \phi\mu^+\mu^-$
decay. In Section 3, we present our numerical analyses and
discussions.
 Section 4 contains our conclusion.

\section{The theoretical framework of  $B_s\to \phi\mu^+\mu^-$ decay}

In the SM, the double differential branching ratio
$\frac{d^2\mathcal{B}}{d\hat{s}d\hat{u}}$ for $B_s\to
\phi\mu^+\mu^-$ may be written as   \cite{Ali:1999mm} {\small
\begin{eqnarray}
\frac{d^2\mathcal{B}^{SM}}{d\hat{s}d\hat{u}}
&=&\tau_B\frac{G^2_F\alpha_{e}^2m_{B_s}^5}{2^{11}\pi^5}|V^*_{ts}V_{tb}|^2\nonumber\\
&&\times\left\{\frac{|A|^2}{4}\Big(\hat{s}(\lambda+\hat{u}^2)+4\hat{m}^2_\mu\lambda\Big)
+\frac{|E|^2}{4}\Big(\hat{s}(\lambda+\hat{u}^2)
-4\hat{m}^2_\mu\lambda\Big)\right.\nonumber\\
&&+\frac{1}{4\hat{m}^2_{\phi}}\Big[|B|^2\Big(\lambda-\hat{u}^2
+8\hat{m}^2_{\phi}(\hat{s}+2\hat{m}^2_\mu)\Big)
+|F|^2\Big(\lambda-\hat{u}^2+8\hat{m}^2_{\phi}(\hat{s}-4\hat{m}^2_\mu)\Big)\Big]\nonumber\\
&&-2\hat{s}\hat{u}\Big[Re(BE^*)+Re(AF^*)\Big]\nonumber\\
&&+\frac{\lambda}{4\hat{m}^2_{\phi}}\Big[|C|^2(\lambda-\hat{u}^2)+
|G|^2(\lambda-\hat{u}^2+4\hat{m}^2_\mu(2+2\hat{m}^2_{\phi}-\hat{s})\Big)\Big]\nonumber\\
&&-\frac{1}{2\hat{m}^2_{\phi}}\Big[Re(BC^*)(1-\hat{m}^2_{\phi}-\hat{s})(\lambda-\hat{u}^2)\nonumber\\
&&~~~~~~~~~+Re(FG^*)\Big((1-\hat{m}^2_{\phi}
-\hat{s})(\lambda-\hat{u}^2)+4\hat{m}^2_\mu\lambda\Big)\Big]\nonumber\\
&&\left.-2\frac{\hat{m}^2_\mu}{\hat{m}^2_{\phi}}
\lambda\Big[Re(FH^*)-Re(GH^*)(1-\hat{m}^2_{\phi})\Big]
+|H|^2\frac{\hat{m}^2_\mu}{\hat{m}^2_{\phi}}\hat{s}\lambda \right\},
\label{BPHI}
\end{eqnarray}
where $ p = p_B+p_{\phi}$, $s = q^2$ and $q = p_++p_-$ ($p_\pm$ the
four-momenta of the muons), and the auxiliary functions $A-H$ can be
found in Ref.  \cite{Ali:1999mm}. The hat denotes normalization in
terms of the B-meson mass, $m_{B_s}$, e.g. $\hat{s}=s/m_{B_{s}}^2$,
$\hat{m}_q=m_q/m_{B_s}$.

In the MSSM without R-parity, the double differential branching
ratio including the squark exchange contribution could be gotten
from Eq. (\ref{BPHI}) by the replacements
 \cite{Xu:2006vk}
{\footnotesize
\begin{eqnarray}
A(\hat{s})&\rightarrow&A(\hat{s})+\frac{1}{W} \left[\frac{2V^{B_s\to
\phi}(\hat{s})}{m_{B_s}+m_{\phi}}m^2_{B_s}\right]\sum_i\frac{\lambda'_{2i2}\lambda_{2i3}'^{*}}{8m^2_{\tilde{u}_{iL}}},\nonumber\\
B(\hat{s})&\rightarrow&B(\hat{s})+\frac{1}{W}
\left[-(m_{B_s}+m_{\phi})A_1^{B_s\to \phi}(\hat{s})\right]
\sum_i\frac{\lambda'_{2i2}\lambda_{2i3}'^{*}}{8m^2_{\tilde{u}_{iL}}},\nonumber\\
C(\hat{s})&\rightarrow&C(\hat{s})+\frac{1}{W}
\left[\frac{A_2^{B_s\to
\phi}(\hat{s})}{m_{B_s}+m_{\phi}}m_{B_s}^2\right]
\sum_i\frac{\lambda'_{2i2}\lambda_{2i3}'^{*}}{8m^2_{\tilde{u}_{iL}}},\nonumber\\
D(\hat{s})&\rightarrow&D(\hat{s})+\frac{1}{W}\left[\frac{~2m_{\phi}}{\hat{s}}\Big(A_3^{B_s\to
\phi}(\hat{s})-A_0^{B_s\to \phi}(\hat{s})\Big)\right]
\sum_i\frac{\lambda'_{2i2}\lambda_{2i3}'^{*}}{8m^2_{\tilde{u}_{iL}}},\nonumber\\
E(\hat{s})&\rightarrow&E(\hat{s})-\frac{1}{W} \left[\frac{2V^{B_s\to
\phi}(\hat{s})}{m_{B_s}+m_{\phi}}m^2_{B_s}\right]\sum_i\frac{\lambda'_{2i2}\lambda_{2i3}'^{*}}{8m^2_{\tilde{u}_{iL}}},\nonumber\\
F(\hat{s})&\rightarrow&F(\hat{s})-\frac{1}{W}
\left[-(m_{B_s}+m_{\phi})A_1^{B_s\to \phi}(\hat{s})\right]
\sum_i\frac{\lambda'_{2i2}\lambda_{2i3}'^{*}}{8m^2_{\tilde{u}_{iL}}},\nonumber\\
G(\hat{s})&\rightarrow&G(\hat{s})-\frac{1}{W}
\left[\frac{A_2^{B_s\to
\phi}(\hat{s})}{m_{B_s}+m_{\phi}}m_{B_s}^2\right]
\sum_i\frac{\lambda'_{2i2}\lambda_{2i3}'^{*}}{8m^2_{\tilde{u}_{iL}}},\nonumber\\
H(\hat{s})&\rightarrow&H(\hat{s})-\frac{1}{W}\left[\frac{~2m_{\phi}}{\hat{s}}\Big(A_3^{B_s\to
\phi}(\hat{s})-A_0^{B_s\to \phi}(\hat{s})\Big)\right]
\sum_i\frac{\lambda'_{2i2}\lambda_{2i3}'^{*}}{8m^2_{\tilde{u}_{iL}}},\nonumber\\
\end{eqnarray}}
where $W=-\frac{G_F\alpha_{e}}{2\sqrt{2}~\pi}V^*_{ts}V_{tb}m_{B_s}$.

The sneutrino exchange contributions are summarized as {\small
\begin{eqnarray}
\frac{d^2\mathcal{B}^{\tilde{\nu}}}{d\hat{s}d\hat{u}}&=&\tau_B\frac{m^3_{B_s}}{2^7\pi^3}
\Bigg\{-\frac{\hat{m}^2_{\mu}}{\hat{m}^2_{\phi}}
\Bigg[Im(WB\mathcal{T}^{*}_{S})
\Big(\lambda^{-\frac{1}{2}}\hat{u}(1-\hat{m}^2_{\phi}-\hat{s})\Big)\nonumber\\
&&+Im(WC\mathcal{T}^{*}_{S})
\lambda^{\frac{1}{2}}\hat{u}-Im(WF\mathcal{T}^{*}_{P})\lambda^{\frac{1}{2}}\nonumber\\
&&+Im(WG\mathcal{T}^{*}_{P})\lambda^{\frac{1}{2}}(1-\hat{m}^2_{\phi})\Bigg]
+|\mathcal{T}_{S}|^2(\hat{s}-2\hat{m}^2_{\mu})\Bigg\},
\end{eqnarray}}
with {\small
\begin{eqnarray}
\mathcal{T}_S=\left[\frac{i}{2}\frac{A_0^{B\to
\phi}(\hat{s})}{\overline{m}_b+\overline{m}_s}
\lambda^{\frac{1}{2}}m^2_{B_s}\right]
\sum_i\left(\frac{\lambda^{*}_{i22}\lambda_{i32}'}{8m^2_{\tilde{\nu}_{iL}}}-\frac{\lambda_{i22}\lambda_{i23}'^{*}}{8m^2_{\tilde{\nu}_{iL}}}\right),\nonumber\\
\mathcal{T}_P=\left[\frac{i}{2}\frac{A_0^{B\to
\phi}(\hat{s})}{\overline{m}_b+\overline{m}_s}\lambda^{\frac{1}{2}}m^2_{B_s}\right]
\sum_i\left(\frac{\lambda^{*}_{i22}\lambda_{i32}'}{8m^2_{\tilde{\nu}_{iL}}}+\frac{\lambda_{i22}\lambda_{i23}'^{*}}{8m^2_{\tilde{\nu}_{iL}}}\right).
\end{eqnarray}}

In the MSSM with R-parity, all the effects arise from the RPC  MIs
contributing to $C_7,\widetilde{C}^{eff}_9,\widetilde{C}_{10}$, and
they are
\begin{eqnarray}
C^{RPC}_7&=&C^{Diag}_7+C^{MI}_7+nC'^{MI}_7,\nonumber\\
 (C^{eff}_{9})^{RPC}&=& (\widetilde{C}^{eff}_{9})^{Diag}+ (\widetilde{C}^{eff}_{9})^{MI}+ n(C'^{eff}_{9})^{MI},\nonumber\\
C^{RPC}_{10}&=&\widetilde{C}^{Diag}_{10}+\widetilde{C}^{MI}_{10}+nC'^{MI}_{10},
\end{eqnarray}
where $n=1$ for the terms related to the form factors $V$ and $T_1$
as well as $n=-1$ for the terms related to the form factors
$A_0,A_1,A_2,T_2$ and $T_3$ in $B_s\to \phi\mu^+\mu^-$ decay.
$C_7^{Diag,MI},(\widetilde{C}^{eff}_{9})^{Diag,MI}$,
$\widetilde{C}^{Diag,MI}_{10}$, $C'^{MI}_7$, $(C'^{eff}_{9})^{MI}$
and $C'^{MI}_{10}$ have been estimated in Refs.
 \cite{Lunghi:1999uk,Cho:1996we,Hewett:1996ct}.  The results for $\mathcal{B}(B_s\to \phi\mu^+\mu^-)$  including MI effects can be obtained from Eq.
(\ref{BPHI}) by the following replacements
 \cite{Altmannshofer:2008dz,Lunghi:2010tr}:
\begin{eqnarray}
C^{SM}_7&\rightarrow&C^{SM}_7+C^{RPC}_7,\nonumber\\
(C^{eff}_{9})^{SM}&\rightarrow& (C^{eff}_{9})^{SM}+ (C^{eff}_{9})^{RPC},\nonumber\\
C^{SM}_{10}&\rightarrow&C^{SM}_{10}+C^{RPC}_{10}.
\end{eqnarray}

From  the double differential branching ratio,  we can get the
dimuon  forward-backward asymmetry
 \cite{Ali:1999mm}
{\small \begin{eqnarray} \mathcal{A}_{FB}(B_s\to \phi\mu^+\mu^-)
=\int d\hat{s}~\frac{\int^{+1}_{-1}\frac{d^2\mathcal{B}(B_s\to
\phi\mu^+\mu^-)}{d\hat{s}dcos\theta}sign(cos\theta)dcos\theta}
{\int^{+1}_{-1}\frac{d^2\mathcal{B}(B_s\to
\phi\mu^+\mu^-)}{d\hat{s}dcos\theta}dcos\theta}.
\end{eqnarray}}

\section{Numerical results and analyses}
In this section, we will investigate the above mentioned physics
observables and study their sensitivity to the new effects due to
the MSSM with and without R-parity. When we study the SUSY effects,
 we consider only one new coupling at one time, neglecting the interferences between
different new couplings, but keeping their interferences with the SM
amplitude. The input parameters are collected in Appendix, and the
following experimental data will be used to constrain parameters of
the relevant new couplings
 \cite{Aaltonen:2011qs,Aaij:2012ac,PDG}
\begin{eqnarray}
&&\mathcal{B}(B_s\to \mu^+\mu^-)<4.5\times10^{-9}  ~(\mbox{at } 95\%~\mbox{CL}),\nonumber\\
&&\mathcal{B}(B \to K\mu^+\mu^-)=(0.48\pm0.06)\times10^{-6}, \nonumber\\
&& \mathcal{B}(B \to
K^*\mu^+\mu^-)=(1.15\pm0.15)\times10^{-6},\nonumber\\
&& \mathcal{B}(B_s \to
\phi\mu^+\mu^-)=(1.47\pm0.52)\times10^{-6}.\label{Eq:exp}
\end{eqnarray}
To be conservative, we use the input parameters varied randomly
within $1\sigma$ error bar and the experimental bounds  at 95\% CL
in our numerical results.

\subsection{The RPV MSSM effects}

Firstly, we  consider the RPV effects in $B_s\to \phi\mu^+\mu^-$
decay. There are three   RPV coupling products, which are
$\lambda'_{2i2}\lambda'^*_{2i3}$ due to squark exchange as well as
$\lambda_{i22}\lambda'^*_{i23}$ and $\lambda^*_{i22}\lambda'_{i32}$
due to sneutrino exchange, relevant to $B_s\to \mu^+\mu^-$, $B_s\to
\phi\mu^+\mu^-$ and $B\to K^{(*)}\mu^+\mu^-$ decays.
We combine the  experimental bounds in Eq. (\ref{Eq:exp}) at 95\% CL
to constrain the three RPV coupling products. Comparing with the
bounds obtained in Ref. \cite{Wang:2011aa},  we find that
$\lambda_{i22}\lambda'^*_{i23}$ and $\lambda^*_{i22}\lambda'_{i32}$
couplings are further constrained by new upper limit of
$\mathcal{B}(B_s\to \mu^+\mu^-)$, and we obtain
$|\lambda_{i22}\lambda'^*_{i23},\lambda^*_{i22}\lambda'_{i32}|\leq1.3\times10^{-4}$.

Using the constrained parameter spaces from the experimental data in
Eq. (\ref{Eq:exp}), we turn to analysis the constrained RPV effects
on the observables of $B_s\to \phi\mu^+\mu^-$ decay which have not
been measured yet.  The s-channel sneutrino exchange couplings
$\lambda_{i22}\lambda'^*_{i23}$ and $\lambda^*_{i22}\lambda'_{i32}$,
which are strongly constrained from $\mathcal{B}(B\to\mu^+\mu^-)$,
 have negligible
contribution to $B_s\to \phi\mu^+\mu^-$ decay. The t-channel squark
exchange coupling $\lambda'_{2i3}
 \lambda'^*_{2i2}$, which is mainly constrained from
 $\mathcal{B}(B\to K^*\mu^+\mu^-)$ and $\mathcal{B}(B_s\to
 \phi\mu^+\mu^-)$, has considerable contribution to $B_s\to \phi\mu^+\mu^-$.
  The effects of the constrained $\lambda'_{2i2}\lambda'^*_{2i3}$ in $B_s\to \phi\mu^+\mu^-$
  are displayed in Fig. \ref{fig:ulplps} by the two-dimensional scatter
plots, and the SM predictions are also shown for comparing
conveniently. The dimuon invariant mass distribution and the dimuon
forward-backward asymmetry are given with vector meson dominance
contribution excluded in terms of $d\mathcal{B}/d\hat{s}$ and
$d\mathcal{A}_{FB}/d\hat{s}$,  and
 included in  $d\mathcal{B}'/d\hat{s}$ and
 $d\mathcal{A}'_{FB}/d\hat{s}$, respectively.
\begin{figure}[th]
\begin{center}
\includegraphics[scale=0.56]{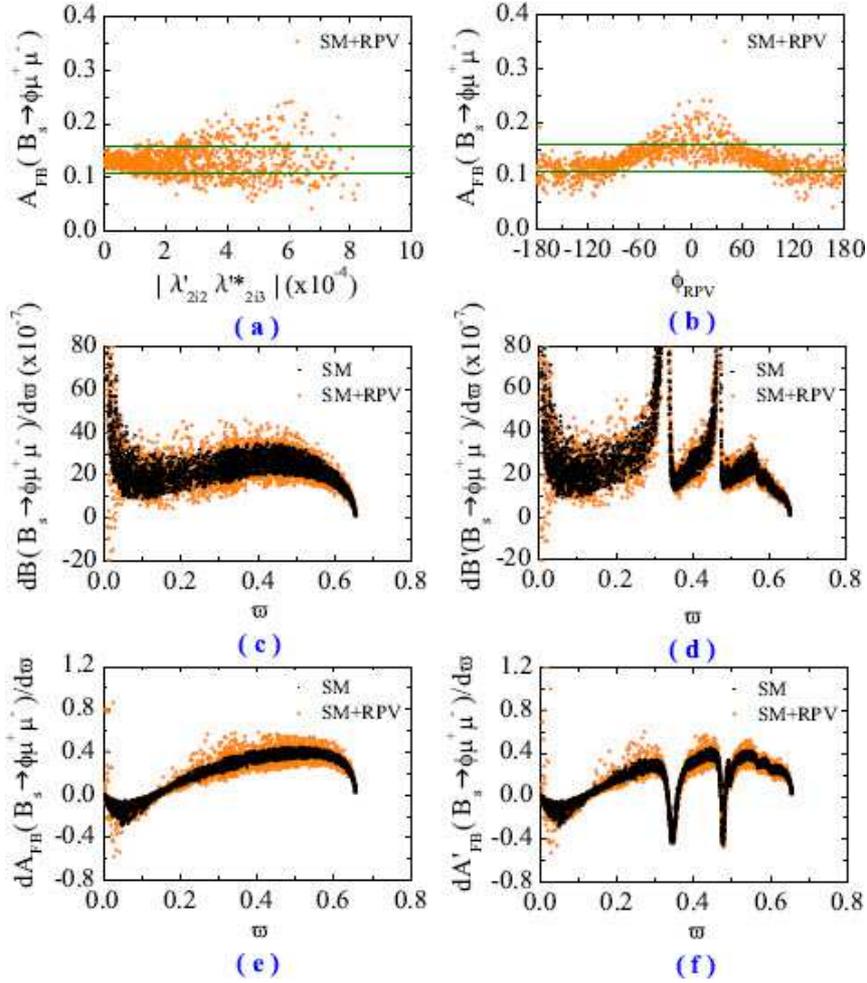}
\end{center}
\vspace{-0.8cm} \caption{The effects of RPV coupling
$\lambda'_{2i2}\lambda'^*_{2i3}$ due to the squark exchange in
$B_s\to \phi\mu^+\mu^-$ decay.  $\phi_{RPV}$ denotes the RPV weak
phase of $\lambda'_{2i2}\lambda'^*_{2i3}$ and $\varpi$ denotes
$\hat{s}$. The limit of SM prediction is shown by olive horizontal
solid lines in plot (a) and (b). }\label{fig:ulplps}
\end{figure}

Now we discuss the plots of Fig. \ref{fig:ulplps} in detail. Figs.
\ref{fig:ulplps} (a) and (b) show the constrained effects of the
modulus and weak phase of $\lambda'_{2i2}\lambda'^*_{2i3}$ on
$\mathcal{A}_{FB}(B_s\to \phi\mu^+\mu^-)$, respectively. One can see
that such contributions could give large $\mathcal{A}_{FB}(B_s\to
\phi\mu^+\mu^-)$  when $|\lambda'_{2i2}\lambda'^*_{2i3}|$ is large
and corresponding $|\phi_{RPV}|$ is near $0^\circ$.
Figs. \ref{fig:ulplps} (c-f) display the constrained RPV effects on
 the dimuon invariant mass spectrum and the differential forward-backward asymmetry,
and we can see that the constrained $\lambda'_{2i2}\lambda'^*_{2i3}$
still has remarkable effects on them. As for the dimuon invariant
mass spectrum, this observable has also been measured as a function
of the dimuon invariant mass square $q^2$ by CDF
\cite{Aaltonen:2011qs}. We do not impose the experimental bound from
$d\mathcal{B}'(B_s\to \phi\mu^+\mu^-)/d\hat{s}$ and leave it as
prediction of the restricted parameter space of
$\lambda'_{2i2}\lambda'^*_{2i3}$, and compare it with the
experimental results in Ref. \cite{Aaltonen:2011qs}.
%
%
The measurement is basically consistent with the SM prediction.
Nevertheless in the region of $2.00<q^2<8.68$ (i.e.
$0.07<\hat{s}<0.31$), the central value of the experimental data
from CDF is smaller than one of its SM predictions. The prediction
of $d\mathcal{B}'(B_s\to \phi\mu^+\mu^-)/d\hat{s}$ including
$\lambda'_{2i2}\lambda'^*_{2i3}$ coupling is allowed by current
experimental data, and the effects of
$\lambda'_{2i2}\lambda'^*_{2i3}$ coupling may be further constrained
if the experimental bound of the dimuon invariant mass spectrum in
Ref. \cite{Aaltonen:2011qs} is considered.

\subsection{The RPC MI effects}

Next, we  explore the RPC MI effects  in $B_s\to \phi\mu^+\mu^-$
decay in the MSSM with large tan$\beta$. There are three kinds of
MIs $(\delta^u_{LL})_{23}$, $(\delta^d_{LL})_{23}$ and
$(\delta^d_{RR})_{23}$ contributing to $B_s\to \mu^+\mu^-$, $B\to
K^{(*)}\mu^+\mu^-$ and $B_s\to \phi\mu^+\mu^-$ decays at the same
time. The experimental data shown in Eq. (\ref{Eq:exp}) will be used
to constrain these three kinds of MI parameters. Our bounds on the
three MI couplings are demonstrated in Fig. \ref{fig:boundMIA}.
Compared with the bounds in Ref. \cite{Wang:2011aa}, the allowed
spaces of all three MI parameters are further constrained by the new
upper limit of $\mathcal{B}(B_s\to \mu^+\mu^-)$.

\begin{figure}[h]
\begin{center}
\includegraphics[scale=0.8]{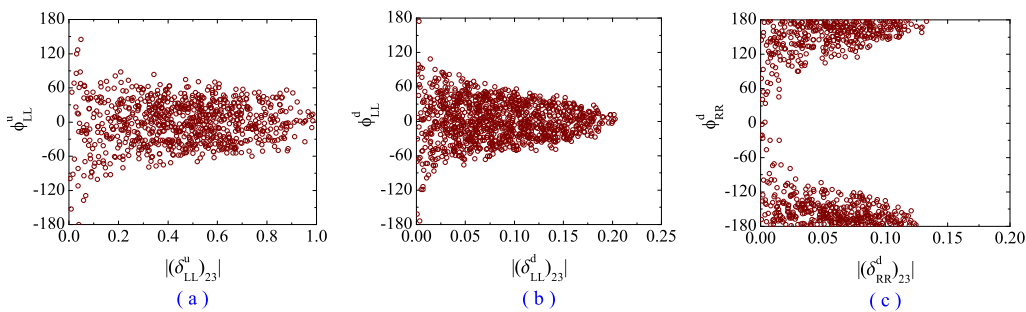}
\end{center}
\vspace{-0.8cm} \caption{The allowed parameter spaces of
$(\delta^u_{LL})_{23}$,  $(\delta^d_{LL})_{23}$ and
$(\delta^d_{RR})_{23}$ MI parameters constrained by
$\mathcal{B}(B_s\to \mu^+\mu^-,\phi \mu^+\mu^-)$ and
$\mathcal{B}(B\to K^{(*)}\mu^+\mu^-)$ at  95\% CL,
 and the RPC phases are given in degree.}\label{fig:boundMIA}
\end{figure}

Then we analyze the RPC supersymmetric effects in $B_s\to
\phi\mu^+\mu^-$ decay. Besides the MI contributions,  the SUSY
predictions  also include the contributions that come from graphs
including SUSY Higgs bosons and sparticles in the limit in which we
neglect all the MI contributions, which are called non-MI
contributions. We find that non-MI couplings  have negligible effect
in $\mathcal{A}_{FB}(B_s\to \phi\mu^+\mu^-)$. The non-MI SUSY
effects on the dimuon invariant mass spectrum and the differential
forward-backward asymmetry of $B_s\to \phi\mu^+\mu^-$ are shown  in
Fig. \ref{fig:nonMI}. As shown in Figs. \ref{fig:nonMI} (a-b),
$d\mathcal{B}(B_s\to \phi\mu^+\mu^-)/d\hat{s}$ could be increased
slightly in the low $\hat{s}$ region, but obviously  decreased  in
the high $\hat{s}$ region.
Figs. \ref{fig:nonMI} (c-d) show us that the non-MI effects could
slightly suppress $d\mathcal{A}_{FB}(B_s\to
\phi\mu^+\mu^-)/d\hat{s}$ at the low $\hat{s}$ ranges.

Since the constrained $(\delta^d_{LL})_{23}$ and
$(\delta^d_{RR})_{23}$ MIs have no obvious effects in $B_s\to
\phi\mu^+\mu^-$, we only show the  $(\delta^u_{LL})_{23}$ MI
contributions to $B_s\to \phi\mu^+\mu^-$ in Fig. \ref{fig:dull}.
Note that the SUSY predictions in Fig. \ref{fig:dull} also include
the non-MI contributions shown in Fig. \ref{fig:nonMI}. From Figs.
\ref{fig:dull} (a-b), one can see that $\mathcal{A}_{FB}(B_s\to
\phi\mu^+\mu^-)$ is very sensitive to $(\delta^u_{LL})_{23}$ MI, and
it increases with $|(\delta^u_{LL})_{23}|$ but decreases with
\begin{figure}[t]
\begin{center}
\includegraphics[scale=0.56]{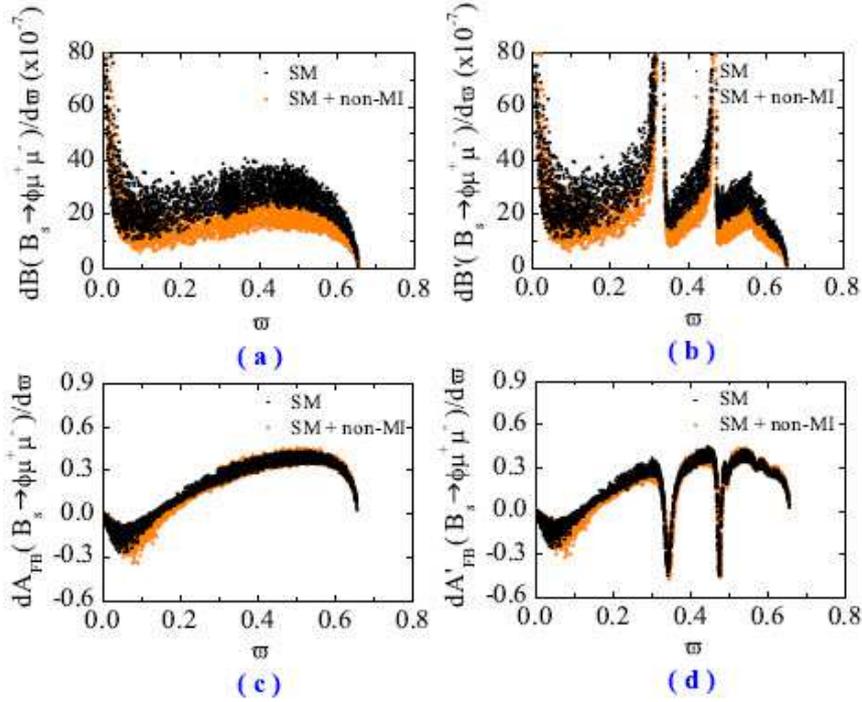}
\end{center}
\vspace{-0.8cm} \caption{The constrained non-MI effects in $B_s\to
\phi\mu^+\mu^-$ decay, and $\varpi$ denotes
$\hat{s}$.}\label{fig:nonMI}
\end{figure}
$|\phi^u_{LL}|$. Figs. \ref{fig:dull} (c-d) show us
$d\mathcal{B}(B_s\to \phi\mu^+\mu^-)/d\hat{s}$ is compatible with
the theoretical uncertainties and thus is indistinguishable from its
SM prediction. As shown in Figs. \ref{fig:dull} (e-f), the
constrained $(\delta^u_{LL})_{23}$ MI effects on
$d\mathcal{A}_{FB}(B_s\to \phi\mu^+\mu^-)/d\hat{s}$ could be
significant. Note that the theoretical uncertainty of
$d\mathcal{A}_{FB}(B_s\to \phi\mu^+\mu^-)/d\hat{s}$ including
$(\delta^u_{LL})_{23}$ MI is smaller than one of
$d\mathcal{A}_{FB}(B\to K^*\mu^+\mu^-)/d\hat{s}$ in Ref.
\cite{Wang:2011aa}.
\begin{figure}[h]
\begin{center}
\includegraphics[scale=0.56]{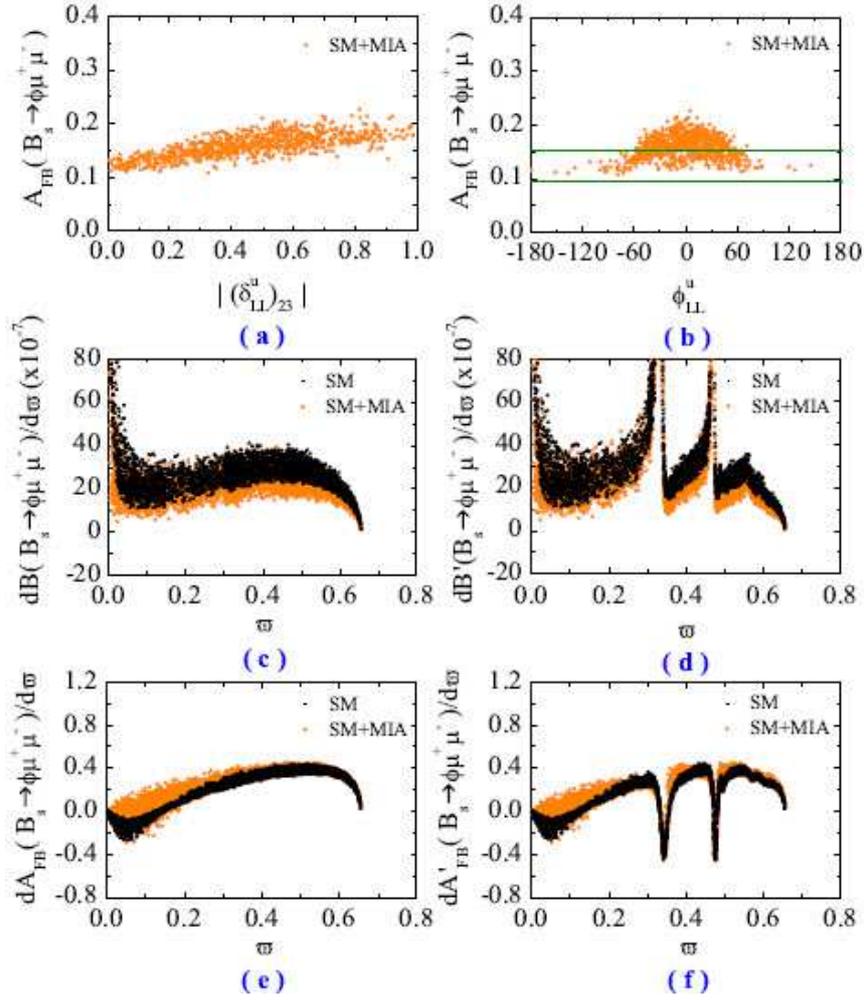}
\end{center}
\vspace{-0.8cm} \caption{The constrained $(\delta^u_{LL})_{23}$ MI
effects in $B_s\to \phi\mu^+\mu^-$ decay, and $\varpi$ denotes
$\hat{s}$.}\label{fig:dull}
\end{figure}

\section{Conclusions}
In this paper,  we have studied $B_s\to \phi\mu^+\mu^-$ decay in the
MSSM with and without R-parity. We have found that the bounds of
sneutrino exchange RPV couplings as well as  $(\delta^u_{LL})_{23}$
and $(\delta^d_{LL,RR})_{23}$ MI couplings  are further constrained
by the new experimental upper limit of $\mathcal{B}(B_s\to
\mu^+\mu^-)$.
The  constrained RPV coupling due to t-channel squark exchange still
has significant effects in $B_s \to \phi\mu^+\mu^-$ decay, and
$\mathcal{A}_{FB}(B_s\to \phi\mu^+\mu^-)$ is sensitive to both the
modulus and the weak phase of this RPV coupling product. The
constrained $(\delta^u_{LL})_{23}$ MI could give large effects on
$\mathcal{A}_{FB}(B_s\to \phi\mu^+\mu^-)$ and
$d\mathcal{A}_{FB}(B_s\to \phi\mu^+\mu^-)/d\hat{s}$ in all $\hat{s}$
region, and besides, $\mathcal{A}_{FB}(B_s\to \phi\mu^+\mu^-)$ is
very sensitive to both $(\delta^u_{LL})_{23}|$ and $(\phi^u_{LL})$,
but the constrained $(\delta^u_{LL})_{23}$ MI has small effects on
$d\mathcal{B}(B_s\to \phi\mu^+\mu^-)/d\hat{s}$.  In addition, the
constrained $(\delta^d_{LL,RR})_{23}$ MIs have ignorable effects on
the observables of $B_s \to \phi\mu^+\mu^-$ decay, nevertheless
$d\mathcal{A}_{FB}(B_s\to \phi\mu^+\mu^-)/d\hat{s}$ could be
distinctly decreased by the SUSY contributions which come from
graphs including SUSY Higgs bosons and sparticles in the limit in
which we neglect all the MI contributions. More precise measurements
at the LHCb and the future super-B factories could test our results
and further shrink or reveal the parameter spaces of MSSM with and
without R-parity.

\section*{Acknowledgments}
The work was supported  by  the National Natural Science Foundation
of China (11105115\&11147136), the Joint Funds of the National
Natural Science Foundation of China (U1204113), the Project of Basic
and Advanced, Technology Research of Henan Province (112300410021),
and the Natural Research Project of Henan Province (2011A140023).

\section*{Appendix: Input parameters}
\label{Appendix} The input parameters are summarized in Table
\ref{INPUT}. For the RPC MI effects, we take the five free
parameters $m_0=450~GeV,m_{1/2}=780~GeV,A_0=-1110,$
$\mbox{sign}(\mu)>0$ and $\mbox{tan}\beta=41$ from Ref.
\cite{Heinemeyer:2012dc}.  All other MSSM parameters are then
determined according to the constrained MSSM scenario as implemented
in the program package SUSPECT \cite{suspect2}.
 For the form factors involving the $B_s\to \phi$
transition, we will use the light-cone  QCD sum rules (LCSRs)
results  \cite{Ball:2004ye,Ball:2004rg},
 which are renewed with  radiative corrections to
the leading twist wave functions and SU(3) breaking effects.
 For the $q^2$ dependence of the form factors,
they can be parameterized in terms of simple formulae with two or
three parameters. The expression can be found in Ref.
 \cite{Ball:2004ye,Ball:2004rg}.  In our numerical data analysis, the
uncertainties induced by $F(0)$  are also considered.

{\footnotesize
\begin{table}[t]
\caption{\small Default values of the input parameters.}
\vspace{0.3cm}\label{INPUT}
\begin{center}
\begin{tabular}{lc}\hline\hline
$m_{B_s}=5.370~GeV,~~m_{B_{u,d}}=5.279~GeV,~~m_W=80.425~GeV,~~m_{\phi}=1.019~GeV,$& \\
$m_{K^\pm}=0.494~GeV,~~m_{K^0}=0.498~GeV,~~m_{K^{*\pm}}=0.892~GeV,~~m_{K^{*0}}=0.896~GeV,$& \\
$\overline{m}_b(\overline{m}_b)=(4.19^{+0.18}_{-0.06})~GeV,~~\overline{m}_s(2GeV)=(0.100^{+0.030}_{-0.020})~GeV,$& \\
$\overline{m}_u(2GeV)=0.0017\sim
0.0031~GeV,~\overline{m}_d(2GeV)=0.0041\sim
0.0057~GeV,$& \\
$m_e=0.511\times10^{-3}~GeV,~~m_\mu=0.106~GeV,$~~
$m_{t,pole}=172.9\pm1.1~GeV. $&  \cite{PDG}\\ \hline
$\tau_{B_s}=(1.466\pm0.059)~ps,~~\tau_{B_{d}}=(1.530\pm
0.009)~ps,~~\tau_{B_{u}}=(1.638\pm 0.011)~ps.$&  \cite{PDG}\\\hline
$|V_{tb}|\approx0.99910,~~|V_{ts}|=0.04161^{+0.00012}_{-0.00078}.$&  \cite{PDG}\\
\hline $\mbox{sin}^2\theta_W=0.22306,~~\alpha_e=1/137.$&
 \cite{PDG}\\\hline $f_{B_s}=0.230\pm0.030~GeV.$&
 \cite{Hashimoto:2004hn}\\\hline\hline
\end{tabular}
\end{center}
\end{table}}
%


\clearpage

\end{document}